\documentclass[
prd
,showpacs ,amssymb,superscriptaddress,aps
]{revtex4}

\usepackage{palatino}
\usepackage{color}

\usepackage{graphicx}
\input{epsf}

\usepackage{amsmath,amssymb}
\usepackage{bm}

\newcommand{\dalm}{\kern1pt\vbox{\hrule height 0.9pt\hbox{\vrule width 0.9pt
\hskip 2.5pt\vbox{\vskip 5.5pt}\hskip 3pt\vrule width 0.3pt}\hrule height 0.3pt}
\kern1pt}

\begin{document}



\title{On the maximum mass limit of neutron stars in scalar-tensor gravity}

\author{Hajime Sotani}
\email{sotani@yukawa.kyoto-u.ac.jp}
\affiliation{Division of Theoretical Astronomy, National Astronomical Observatory of Japan, 2-21-1 Osawa, Mitaka, Tokyo 181-8588, Japan}

\author{Kostas D. Kokkotas}
\affiliation{Theoretical Astrophysics, IAAT, Eberhard-Karls University of T\"{u}bingen, 72076 T\"{u}bingen, Germany}
\affiliation{Division of Theoretical Astronomy, National Astronomical Observatory of Japan, 2-21-1 Osawa, Mitaka, Tokyo 181-8588, Japan}

\date{\today}

\begin{abstract}
The maximum mass limits of neutron stars in scalar-tensor gravity is discussed and compared with the limits set by general relativity. The limit is parametrized with respect to the combination of the nuclear saturation parameters and the maximum sound velocity in the core. 
It is shown that, for smaller values of the sound velocity in the core, the maximum mass limit of the scalarized neutron stars is larger than that  of in general relativity. However, for stiff equations of state with sound velocity  higher than $79\%$ of the velocity of light, the maximum mass limit  in general relativity is  larger than that of in scalar-tensor gravity. The results suggest that future observations of massive neutron stars, may constrain the maximum sound velocity as well as the coupling parameter in scalar-tensor gravity.
\end{abstract}

\pacs{04.40.Dg, 04.50.Kd, 04.80.Cc}
%
\maketitle
\section{Introduction}
\label{sec:I}

General relativity (GR) is the prevailing theory for gravity. For hundred years has been tested in various astrophysical scenarios and its agreement with experiments and observations is remarkable. 
Still, most of GR's tests  have been performed  in the weak-field regime, e.g. in our solar system.
Moreover, there exist phenomena, such as dark matter and dark energy (accelerated expansion of the universe) which do no fit well in the standard GR framework.
At smaller scales, it is not yet clear if gravitational field is described  by GR in the strong-field regime, for example around/inside a compact object. Thus, one might probe and/or verify the gravitational theory in strong-field regime via the astronomical observations of the phenomena associated with compact objects. In fact, several possibilities for distinguishing observationally the alternative theories from general relativity were suggested \cite{DP2003,SK2004,S2014a,S2014b,B2015}.

Among the many proposed possibilities for testing the theory of gravity and understanding the physics of matter in the high-density region the bulk properties  of relativistic stars (mass and radius) emerge as the top candidates. 
Actually, due to the nuclear saturation, the properties of nuclear matter in the high density region cannot fully constrained via the terrestrial nuclear experiments. As a result, several equations of state (EOSs) for neutron star matter have been proposed theoretically. However, the discoveries of $2M_\odot$ neutron stars ruled out most of the soft EOSs, for which the maximum mass can not reach  the observed $2M_\odot$ \cite{D2010,A2013}. These observations led also to the so called ``hyperon puzzle", see \cite{vanDal2014, Fortin2016,Bom2016}. 
Detail studies in GR, demonstrated that the maximum mass of neutron stars depends strongly on the stiffness of the EOS  \cite{H1978,KSF1997,Sotani2016}.

One of the most natural generalizations of GR is to include in the field equations an additional mediator -- a scalar field $\Phi$. These are the so-called Scalar-Tensor theories (STT) originated  in the 1950-60's with the works of Jordan, Fierz, Brans and Dicke \cite{Jordan49,Fierz56,Jordan59,Brans61,Dicke62}. The study of compact stars in STT revealed a new phenomenon, that is,  the so-called spontaneous scalarization  \cite{DE1993,Harada1998}. If a compact star is subject to scalarization its bulk properties can significantly deviate from those in general relativity. This phenomenon typically happens in the more compact models of neutron stars. In this case, the central density of the scalarized neutron star ($\tilde{\rho}_c$) should be relatively large such as $\tilde{\rho}_c \gtrsim 2\tilde{\rho}_0$, where $\tilde{\rho}_0$ denotes the saturation density. 

In this article we examine the maximum mass limit of scalarized neutron stars in scalar-tensor theory of gravity. 
For the neutron star models,  we adopt a ``phenomenological EOS'' for the description of the low-density region ($\rho \lesssim 2\tilde{\rho}_0$),  proposed  by Oyamatsu and Iida \cite{OI2003,OI2007}, which characterised solely by the nuclear saturation parameters \cite{SIOO2014,SSB2016}.
In the high-density region, there are many uncertainties regarding the appropriate EOS. Thus, we adopt a simple EOS, parametrized by the maximum sound velocity directly associated with the stiffness of the EOS, as in \cite{H1978,KB1996,Sotani2016,ST1983}. The EOS for the low and high density are matched at  $\tilde{\rho}=2\tilde{\rho}_0$. 

In the next two sections (\ref{sec:II} and \ref{sec:III}), we discuss the details of the scalar-tensor theory used in this work and we review the current understanding on the equations of state. In section \ref{sec:IV}, we present the results for the maximum mass of neutron stars in scalar-tensor theory of gravity and we demonstrate its dependence on the sound velocity, the combination of the nuclear saturation parameters and the coupling constant of the scaler field. The results are discussed in section \ref{sec:V}.

 We adopt geometric units, $c = G_* = 1$, where $c$ and $G_*$ denote the velocity of light and the gravitational constant, respectively.

\section{Neutron stars in Scalar-tensor theory of gravity}
\label{sec:II}

The form of the action describing scalar-tensor theory of gravity with one scalar field, $\varphi$, in the ``Einstein frame''  is given by
\begin{equation}
   S = \frac{1}{16\pi G_*}\int\sqrt{-g_*}\left(R_* -2g_*^{\mu\nu}\varphi_{,\mu}\varphi_{,\nu}\right)d^4x
      + S_m\left[\Psi_m, {\cal A}^2g_{*\mu\nu}\right].
\end{equation}
Here the quantities with asterisk are defined in the frame associated with the ``Einstein metric" $g_{*\mu\nu}$. That is, $R_*$ is the curvature scalar constructed from $g_{*\mu\nu}$ and $g_*$ is the determinant of $g_{*\mu\nu}$, while $G_*$ is the bare gravitational coupling constant. $S_m$ is the action for the matter represented by $\Psi_m$. Actually,  $\Psi_m$ expresses all matter fields collectively. 
The field equations in the Einstein frame become simpler than those in the ``Jordan'' or ``physical frame''. Thus, one may work on the Einstein frame but compares the theoretical results  with experimental or observational data derived/collected in the Jordan frame.  
%
The ``Jordan metric" $\tilde{g}_{\mu\nu}$,  is related to the ``Einstein metric'' via the conformal transformation, 
\begin{equation}
  \tilde{g}_{\mu\nu} = {\cal A}^2(\varphi)g_{*\mu\nu}.
\end{equation}
Hereafter, the tilded quantities and quantities with asterisk will be respectively considered as the variables in the Jordan frame and in the Einstein frame.

The variation of the action $S$ with respect to $g_{*\mu\nu}$ and $\varphi$, leads to the following set of field equations in the Einstein frame
\begin{eqnarray}
  G_{*\mu\nu} &=& 8\pi G_*T_{*\mu\nu} + 2\left(\varphi_{,\mu}\varphi_{,\nu} - \frac{1}{2}g_{*\mu\nu}g_*^{\alpha\beta}\varphi_{,\alpha}\varphi_{,\beta}\right), \label{eq:field1}  \\
  \dalm_* \varphi &=& -4\pi G_* \alpha(\varphi)T_*, \label{eq:field2}
\end{eqnarray}
where $T_{*\mu\nu}$ is the energy-momentum tensor in the Einstein frame. $T_{*\mu\nu}$ is related to the physical energy-momentum tensor $\tilde{T}_{\mu\nu}$ via the relation 
\begin{equation}
  T_*^{\mu\nu} \equiv \frac{2}{\sqrt{-g_*}}\frac{\partial S_m}{\partial g_{*\mu\nu}} = {\cal A}^6\tilde{T}^{\mu\nu} \, ,
\end{equation}
while $T_*$ and $\alpha(\varphi)$ are given by 
\begin{eqnarray}
  T_*  &\equiv& T_{*\mu}^{\ \ \mu} = T_*^{\mu\nu} g_{*\mu\nu} \, , \\
  \alpha(\varphi) & \equiv & \frac{d \ln {\cal A}(\varphi)}{d\varphi}.
\end{eqnarray}
In addition to the field equations (\ref{eq:field1}) and (\ref{eq:field2}), the energy-momentum conservation  is expressed as $\tilde{\nabla}_\nu\tilde{T}_{\mu}^{\ \nu}=0$ in the Jordan frame, or as
\begin{equation}
   \nabla_{*\nu}T_{*\mu}^{\ \ \nu} = \alpha(\varphi)T_*\nabla_{*\mu}\varphi
\end{equation}  
in the Einstein frame. This version of scalar-tensor  theory for gravity in the limit $\alpha=0$ reduces to general relativity.

For the conformal factor ${\cal A}(\varphi)$, we adopt the same functional form as in Damour and Esposito-Far\`{e}se \cite{DE1993}, i.e.,
\begin{equation}
   {\cal A}(\varphi) = \exp\left(\frac{1}{2}\beta\varphi^2\right),
\end{equation}
which leads to that $\alpha=\beta\varphi$. The coupling constant $\beta$ is a real number, and  the spontaneous scalarization sets in for $\beta \lesssim -4.35$ \cite{Harada1998}. Notice that for $\beta=0$, in this formulation, the scalar-tensor theory reduces to general relativity. For fast rotating neutron stars the spontaneous  scalarization sets in for larger values of $\beta$. Actually, for rotating neutron stars in the break up or Kepler limit scalarization sets in for $\beta\lesssim -3.9$ \cite{DYSK2013}. The so-called ``dynamical scalarization'' may take place during the post-merger phase of binary neutron star systems as suggested in \cite{PBPL2014,TSB2015}. In this case the scalarization may sets in for larger values of  $\beta$. For such high values of  $\beta$ the equilibrium (or static) spontaneous scalarization found for spherically symmetric compact objects does not sets in.  Actually, the constraints to $\beta$ set by observations are quite severe. Observational data, from binary neutron star - white dwarf systems  \cite{Freire2012}, suggest that  $\beta \gtrsim -5$. 

Typically, the sequences of neutron star models are constructed by integrating the modified Tolman-Oppenheimer-Volkoff equations assuming  an appropriate EOS \cite{Harada1998,SK2004}. In order to define the scalar field distribution, one has to choose an asymptotic value $\varphi_0$ for the scalar field,  here we assume $\varphi_0=0$. Then, one may construct a one parameter (central density $\rho_c$) family of neutron star models by fixing the EOS and the value of $\beta$. 

In the next section, we described the parametrized EOS  adopted in this paper, for the study of the maximum mass limit of neutron stars in scalar-tensor gravity.

\section{Nuclear saturation parameters and EOS}
\label{sec:III}

As discussed in Lattimer \cite{L1981,LP2016}, the bulk energy of nuclear matter at zero temperature can be expanded as a function of the baryon number density and the neutron excess, where the coefficients of the expansion are the nuclear saturation parameters. Actually, any EOSs can be expanded in a similar way, and as a result to be characterized by a unique set of nuclear saturation parameters. Several terrestrial nuclear experiments, set constraints in the values of the nuclear saturation parameters (see the discussion in Lattimer and Lim \cite{LL2013}). Even so, the incompressibility parameter of symmetric nuclear matter, $K_0$,
and the density dependent nuclear symmetry energy, the so-called slope parameter $L$, 
are relatively more difficult to constrain from the experiments \cite{Newton2014}.

Here, we assume that the nuclear properties for densities lower than $\sim 2\tilde{\rho}_0$ are expressed solely via the nuclear saturation parameters. In fact, the neutron star models in general relativity with the central density in the range of $\tilde{\rho}_c\lesssim 2\tilde{\rho}_0$ can be described nicely with the combination of $K_0$ and $L$ via $\eta=(K_0 L^2)^{1/3}$ \cite{SIOO2014,SSB2016}. Thus, for the lower density region, i.e., for $\tilde{\rho}\lesssim 2\tilde{\rho}_0$, we adopt the ``phenomenological'' EOS proposed by Oyamatsu and Iida \cite{OI2003,OI2007}. In this approach,  the EOS is constructed for values of $K_0$ and $L$ that are in agreement with the experimental data \cite{OI-EOS}. Here, we adopt the same sets of the nuclear saturation parameters as in Ref. \cite{Sotani2016}. Actually, the values of $K_0$ and $L$ obtained from the terrestrial experiments are $K_0=230\pm 40$ MeV \cite{Khan2013} and $30\lesssim L\lesssim 80$ MeV \cite{Newton2014}, respectively. These experimental constraints suggest that the allowable values for $\eta$ lie in the range $55.5\lesssim \eta\lesssim 120$ MeV.

The composition of dense nuclear matter in the high density (deep interior) of neutron stars is still largely unknown. A better understanding of the composition of nuclear matter at the highest densities has important implications not only in nuclear physics, but also in relativistic astrophysics, as it is connected to the conditions for equilibrium and the onset of dynamical and secular instabilities of neutron stars.  In the last fifty years a number of EOSs have been proposed, which are based on the various nuclear interactions, nuclear theory, and compositions. A fundamental upper limit for the high density region of neutron stars comes from the demand that the sound velocity should be smaller than unity (velocity of light). An obvious lower limit can be set by the condition  that the sound velocity should be larger than zero for thermodynamical stability. 

An approximate parametrization of the stiffest EOS 
is given by $\tilde{p}=\tilde{p}_t+\tilde{\rho}-\tilde{\rho}_t$. Here $\tilde{p}_t$ is the pressure at the transition density,
 $\tilde{\rho}_t$,  determined from the well understood EOS in low density region. Moreover, it was conjectured that the maximum sound velocity should be less than $c/\sqrt{3}$ \cite{BS2015}. This assumption was used for the study of the  neutron star maximum mass limit assuming, for the higher density region, a parametrization for the  EOS of the form $\tilde{p}=\tilde{p}_t+(\tilde{\rho}-\tilde{\rho}_t)/3$. Here we adopt a similar parametrization in the study of  the maximum mass of neutron stars in scalar-tensor gravity. More specifically we assume : 
\begin{equation}
  \tilde{p} = \tilde{p}_t + \bar{v}_s^2 (\tilde{\rho} - \tilde{\rho}_t),
\end{equation}
where $\bar{v}_s$ denotes the maximum sound velocity in the high density region of the neutron star. This parametrized high-density EOS is matched to the ``phenomenological'' EOS for the lower density region at $\tilde{\rho}_t=2\tilde{\rho}_0$. Furthermore, we adopt a quite wide range of values for  the sound velocity, more specifically we consider $1/3\le \bar{v}_s^2\le 1$. 

The aforementioned assumptions lead to a parametrized study of the maximum mass limit of neutron stars in scalar-tensor theory. More specifically the low density regime will be parametrized by $\eta$, the high density regime by $\bar{v}_s$ while the effect of the scalar field will enter via the coupling-parameter $\beta$.

\section{The maximum mass limit of neutron stars in scalar-tensor gravity}
\label{sec:IV}

\begin{figure*}
\begin{center}
\includegraphics[scale=0.5]{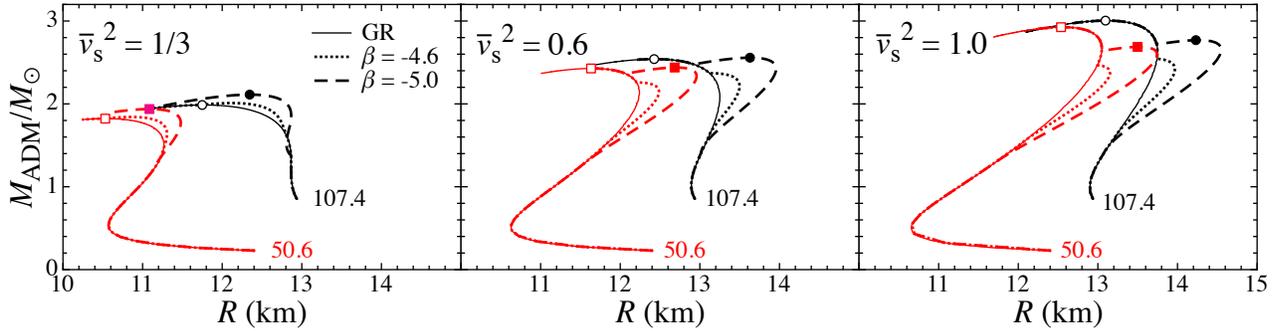}
\end{center}
\caption{
ADM mass versus radius diagrams for neutron stars in GR and STT. The sequences of neutron star models generated for two EOS, in the low density regime, assuming $\eta=50.6$ and 107.4 MeV. The solid lines denote the mass-radius relation in GR, while the dotted and dashed lines denote the mass-radius relation  in STT for $\beta=-4.6$ and $-5.0$. The three panels from left to the right  correspond to sound velocity values $\bar{v}_s^2=1/3$, 0.6, and 1.  The open (filled) marks show the maximum mass limit of neutron star in GR (in STT for   $\beta=-5.0$).
}
\label{fig:MR}
\end{figure*}

 We studied two characteristic sequences of neutron star models in general relativity and in scalar-tensor theory of gravity using the parametrization described earlier. More specifically, the two sequences, of ADM mass versus radius, are presented in Fig. \ref{fig:MR},  where the parametrization was done with respect to $\eta$.
The solid lines correspond to general relativistic results while the dotted and dashed lines correspond to sequences of neutron stars in scalar tensor theory with $\beta=-4.6$ and $-5.0$ correspondingly. The three panels, parametrized with respect to the sound velocity, which was chosen to be $\bar{v}_s^2=1/3$, 0.6, and 1.0. Finally, the filled end empty marks correspond to the maximum mass models of the sequences.
It should be noted, that the stellar models in the STT are the identical to the  GR ones for the smaller mass models. In the models of Fig. 1 scalarization takes place for masses $\sim1.4 M_\odot$. The three sequences tend to merge again for  models with larger masses. In the later case, the more massive stellar models in  STT are approaching those of  GR and scalarization does not takes place. 
 
In the left panel of Fig. \ref{fig:MR}, we observe  that  for small values of the sound velocity the maximum mass limit of the scalarized neutron stars is larger than the corresponding limit in general relativity. This is a quite well known result from earlier studies \cite{Harada1998,S2014a}.  As the value of the sound velocity increases the maximum mass limit in GR increases faster and for $\bar{v}_s^2 \approx 0.6$ the two sequences have the same limit in both theories. Moreover, for larger values of the sound velocity the general relativistic sequence has a larger maximum mass limit.  Actually, from  this last case, one may draw an additional conclusion. That is, if the sound velocity is extremely high then even if the true theory for gravity is the STT, the neutron star models with the maximum mass will not be scalarized, and the maximum mass limit will coincide with the prediction of general relativity.

\begin{figure}
\begin{center}
\includegraphics[scale=0.5]{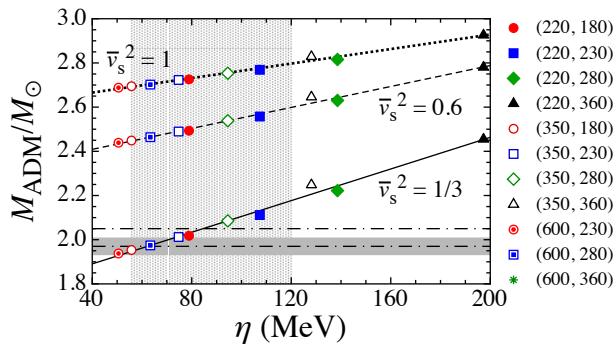}
\end{center}
\caption{
The dependence of the maximum mass of scalarized neutron stars as function of the parameter $\eta$ is shown.  Three sequences for characteristic values of the sound velocity and fixed $\beta$ ($=-5.0$) are plotted.  The various marks correspond to different combinations of the nuclear saturation parameters $(-y,K_0)$ \cite{OI-EOS}. The solid, broken, and dotted lines are linear fittings given by Eq. (\ref{eq:M-eta}) for $\bar{v}_s^2=1/3$, 0.6, and 1, respectively.
The stippled region denotes the plausible range  of values for $\eta$ as constrained from the terrestrial nuclear experiments. For reference, we also draw the the values of the masses of the two most massive neutron stars known to exists i.e. the  PSR J1614-2230 \cite{D2010} and  PSR J0348+0432 \cite{A2013}.
}
\label{fig:eta-M-b50}
\end{figure}

In general relativity, the dependence of the maximum mass  on the parameter $\eta$, was extensively discussed in \cite{Sotani2016}, were also  fitting formulae have been derived. Here, we extent this study by taking into account the  dependence of the maximum mass limit on $\eta$ for scalarized neutron star models. 
More specifically, the results for fixed value of the coupling parameter, $\beta=-5.0$, for scalarized neutron stars are presented in 
Fig. \ref{fig:eta-M-b50}. Three sequences of scalarized neutron star models were generated. They were parametrized by the sound velocity, $\bar{v}_s^2$ (=1/3, 0.6, and 1.0). It is quite remarkable that all three sequences demonstrate that the ADM mass depends linearly on the parameter $\eta$. Interestingly, the maximum mass dependents solely on $\eta$ and not on the individual values of the nuclear saturation parameters $y$ and $K_0$ \cite{OI-EOS}. In Fig. \ref{fig:eta-M-b50} the plausible range for $\eta$ ($55.5\lesssim \eta\lesssim 120$ MeV) is shown with the stippled region. 

The linear fitting of the maximum ADM mass as function of $\eta$ is of the form:
\begin{equation}
  \frac{M_{\rm ADM}}{M_\odot} = a_1 + a_2 \left(\frac{\eta}{1\ {\rm MeV}}\right),   \label{eq:M-eta}
\end{equation}
where $a_1$ and $a_2$ are appropriate coefficients depending on $\bar{v}_s^2$ and $\beta$. 
Furthermore, we find that the numerical results can be easily fitted via the following relations
\begin{eqnarray}
  a_1(\bar{v}_s^2) &=& 3.0254-0.4263\, \bar{v}_s^{-2}\, ,\label{eq:a1} \\
  a_2(\bar{v}_s^2) &=& \left(0.8183\,\bar{v}_s^{-2} + 1.2946 - 0.4771 \bar{v}_s^2 \right) \times 10^{-3}.  \label{eq:a2}
\end{eqnarray}

\begin{figure}
\begin{center}
\includegraphics[scale=0.5]{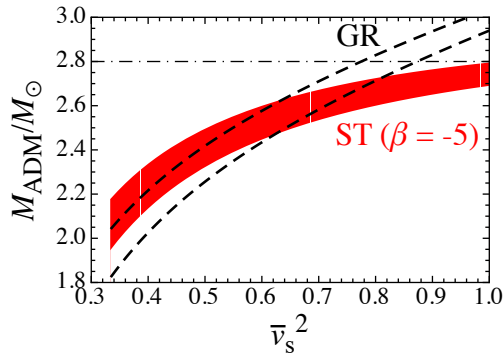}
\end{center}
\caption{
The range of maximum mass limits of neutron stars in general relativity (the region between the broken lines) and of scalarized neutron stars in scalar-tensor gravity with $\beta=-5.0$ (the shaded region) as a function of $\bar{v}_s^2$. Here for $\eta$ we adopted the plausible values $55.5 \lesssim \eta \lesssim 120$ MeV.
}
\label{fig:vs2-M-b50}
\end{figure}

The fitting formulae  (\ref{eq:M-eta}) -- (\ref{eq:a2})  lead to a unique parametrization of maximum neutron star mass in scalar-tensor theory for a given coupling constant $\beta$ (here we assumed $\beta=-5$).  This parametrization is depicted in Fig. \ref{fig:vs2-M-b50}. More, specifically in Fig. \ref{fig:vs2-M-b50} the maximum mass limit  is plotted as function of the sound velocity for the plausible range of values for $\eta$ ($=55.5$ -- 120 MeV) described earlier. In this figure the main feature presented in Fig. \ref{fig:MR} are verified. More specifically, for small sound velocity values scalar-tensor theory allows for larger maximum mass models. As the value of the sound velocity increases  general relativity allows for larger maximum mass models than scalar-tensor theory. Actually, there is set of critical values for $\bar{v}_s^2$ and $\eta$ where the transition happens. For the specific neutron star sequences  i.e. for $\beta=-5$, the critical value of $\bar{v}_s^2$ where the maximum mass of scalarized neutron stars becomes smaller than that of neutron stars in general relativity, is $\bar{v}_s^2=0.62$ for $\eta= 55.5$ MeV and $\bar{v}_s^2=0.64$ for $\eta=120$ MeV. These values correspond to 79 \% and 80 \% of velocity of light in vacuum. 

The results in Fig. \ref{fig:vs2-M-b50} show that, in the limit $\bar{v}_s \to 1$, the  maximum neutron star mass in scalar-tensor gravity, for $\beta=-5.0$ and for plausible values of $\eta$, is $\simeq 2.8M_\odot$.
This upper limit suggests, that if non-rotating or slowly rotating neutron stars observed with masses exceeding this limit, they will be non-scalarized even if the scalar-tensor gravity for $\beta\ge -5$ is the correct theory of gravity in strong-field regime.

As it was shown in \cite{Sotani2016}, in the general relativistic framework  there is a difficulty in explaining the observed masses of PSR J1614-2230 and PSR J0348+0432 under the assumption that the  maximum sound velocity in the core is smaller  than $c/\sqrt{3}$ \cite{BS2015}. However, at this range of  sound velocity values,  scalarization allows for neutron stars with masses even larger than the observed ones.  In fact, by considering the scalar-tensor gravity with $\beta=-5.0$, one may explain the lower mass limit constrained by observations for PSR J0348+0432 for a wide range of plausible values of $\eta$. On the contrary,  if one considers more massive neutron stars, such as J1748-2021B  with $M = (2.74 \pm 0.21)M_\odot$ \cite{Freire2008}, then even if scalar-tensor gravity is the correct theory of gravity in strong-field regime, such an object is will not be scalarized.  Such an observation  will, set, in addition,  an upper limit for the possible maximum sound velocity $\bar{v}_s$. According to  Fig. \ref{fig:vs2-M-b50} the critical sound velocity should be $\bar{v}_s^2\gtrsim 0.77$, which corresponds to the intersection of the upper broken line and $M_{\rm ADM}\simeq 2.8M_\odot$. 

\section{Conclusion}
\label{sec:V}

 We examined the maximum mass limit of neutron stars in scalar-tensor gravity. A phenomenological EOS was adopted for the lower density part of the stars and a stiffest EOS for the higher density region. The two EOS were matched at the point where the density was twice the saturation one. The low density  EOSs was parametrized with respect to a combination of the nuclear saturation parameters  ($\eta$) while the EOS for the high density was parametrized with respect to the maximum sound velocity ($\bar{v}_s$). Under these assumptions and for fixed value of the coupling constant $\beta$ empirical relations were derived for the maximum mass of the neutron stars in scalar tensor theory.  The empirical formulae express in a unique way the maximum ADM mass limit as functions of the two parameters of the study i.e.  $\eta$ and $\bar{v}_s^2$. 
 
 The study shows that, for lower values of $\bar{v}_s^2$,  the maximum mass limit of scalarized neutron stars is larger than that of neutron stars in general relativity. As the sound velocity increases and especially for values larger than the critical value $\bar{v}_s^2\approx 0.62-0.64$ the maximum mass limit of the scalarized neutron stars is lower than that of general relativity. This shows that even if in the strong field regime, scalar-tensor theory is the true theory of gravity,  for the stiffest of the  EOS, the neutron stars will not be scalarized.
 
 Moreover, the maximum mass allowed by scalar-tensor theory was also derived.  Since observations suggest  $\beta\ge-5.0$ and the maximum mass decreases as  $\beta$ increases the limits set here for the maximum mass of scalarized stars could be considered as the maximum mass allowed by the theory. 
 
Another conclusion that maybe drawn from  Fig. \ref{fig:MR} is that not only the mass but also the radius depends strongly on $\bar{v}_s^2$ both in GR and in STT. Actually, the dependence of radius on $\bar{v}_s^2$ is even more pronounce. This potentially may affect  the constraints that will be set by observations on the radius and the EOS  if the correct theory of gravity is not taken into account.  Thus future attempts in constraining the EOS should consider the possibility that  GR may not be the prevailing theory for gravity. On the other hand further constraints on $\beta$ from independent observations might reduce even further the parameter space for which scalarization is possible.  Still, as it was shown in \cite{Yazadjiev2014}, for fast-rotating neutron stars, the scalarization is possible for a wider range of values of $\beta$. As a result, even though the non-rotating models of an EOS are not scalarized  or the scalarization is so small that it is not observable the rotating models of the same EOS will be scalarized for a certain range of values of $\beta$ and of the central density. This leads to the following scenario, if the scalarization is not important or pronounced for non-rotating stars then one may use the observations of slowly rotating neutron stars to constrain the EOS and then test the theory of gravity via the fast rotating models of this EOS. This calls for extension of the present study to fast rotating scalarized neutron star models.


 The approach used here can be applied in the study of neutron stars in other alternative theories of gravity, e.g. variations of $F(R)$ gravity
 \cite{Yazadjiev2014,Sultana2014,Capozziello2016,AO2016}.
 In all these theories the maximum mass limit is expected to depend in a similar fashion on $\eta$, $\bar{v}_s^2$, and the coupling parameter(s) of the theory.

\acknowledgments
We are grateful to K. Oyamatsu and K. Iida for preparing the EOS table. This work was supported in part by Grant-in-Aid for Young Scientists (B) through Grant No. 26800133 provided by Japan Society for the Promotion of Science (JSPS) and by Grants-in-Aid for Scientific Research on Innovative Areas through Grant No. 15H00843 provided by MEXT.  Part of this work was support by a Visiting Scholar Program grant awarded by the Research Coordination Committee, National Astronomical Observatory of Japan (NAOJ).



\end{document}